# INTERACTIVE GIS WEB-ATLAS
# FOR TWELVE PACIFIC ISLANDS COUNTRIES

**Fabrice LARTIGOU, Michael GOVOROV, Tofiga AISAKE
and Pankajeshwara N. SHARMA**
GIS Unit, Department of Geography and Department of Mathematics and Computer Science,
The University of the South Pacific, P.O. Box 1168, Suva, Fiji Islands,
Fax: (679) 330 1487, E-mails: lartigou_f@usp.ac.fj & govorov_m@usp.ac.fj

**Abstract**

This article deals with the development of an interactive up-to-date *Pacific Islands Web GIS Atlas*. It focuses on the compilation of spatial data from the twelve member countries of the University of the South Pacific (Cook Islands, Fiji Islands, Kiribati Islands, Marshall Islands, Nauru, Niue, Tonga, Tuvalu, Tokelau, Solomon Islands, Vanuatu, and Western Samoa). A previous bitmap web Atlas was created in 1996, and was a pilot activity investigating the potential for using Geographical Information Systems (GIS) in the South Pacific. The objective of the new atlas is to provide sets of spatial and attributive data and maps for use of educators, students, researchers, policy makers and other relevant user groups and the public.

GIS is a highly flexible and dynamic technology that allows the construction and analysis of maps and data sets from a variety of sources and formats. Nowadays, GIS application has moved from local and client-server applications to a three-tier architecture: Client (Web Browser) – Application Web Map Server – Spatial Data Warehouses. The objective of this project is to produce an Atlas that will include interactive maps and data on an Application Web Map Server.

Intergraph products such as GeoMedia Professional, Web Map and Web Publisher have been selected for the web atlas production and design. In an interactive environment, an atlas will be composed from a series of maps and data profiles, which will be based on legend entries, queries, hot spots and cartographic tools. Only the first stage of development of the atlas and related technological solutions are outlined in this article.

**Introduction**

Today, Computer Science and GIS technologies are leading to the improvement of research and technology in the fields of interactive and dynamic Web-Map services and on-line geo-spatial data processing. The recent development of web-map projects in the global GIS users' community is mostly located in developed countries, with little application to the unique and often smaller-scale social, economical, and environmental requirements of the developing world community.

Given the size of the South Pacific, it makes sense that GIS would be an attractive tool for use in the management of land and sea resources. Public and private sector organizations in the South Pacific have been building such systems for a little over a decade. The development of a strong GIS user community requires a broad base for GIS training and education within the region. The University of the South Pacific (USP) is a regional university owned collectively by the governments of the twelve countries.

The USP GIS and Remote Sensing program has experienced considerable growth over the past ten years, and is currently serving an increasingly large GIS user community. The Pacific Web Atlas is considered be one of the main development factors in the Geographical education in this region. A first bitmap Web Atlas was created in 1996, edited by Dr. Bruce Davis. The first atlas design was created in HTML. It presented basic general-reference and thematic maps of the counties of the region and large islands, graphs and major social-economical characteristics (population, various measures of economic activity and well-being). Another major development in Atlas cartography in the South Pacific was the publication of *An Atlas of Fiji* in 1998, edited by USP Deputy Vice-Chancellor Prof. Rajesh Chandra.

During the last eight years, the 1996 Web Atlas was intensively used; many people visited the GIS Unit web site and sent their appreciations. However, changing technology and the availability of new sets of digital data show that there is an urgent need to update the Atlas. A new Pacific Islands Web-Atlas project was proposed, based on the Web mapping technology of Application Server, publication of detailed and up-to-date spatial and attributive data and broadening of the atlas content.

The objectives of the new Pacific Islands Web Atlas project are to develop a methodology in Web-map design for the Pacific Islands Atlas. One of the challenges of the research and methodology development will be to find the right scale ranges for the cartographic representations of twelve very different USP regional countries. Research of spatial data from each USP member countries and examination of the spatial data availability in each country will be one of the key elements for the web-atlas success. At the same time, issues of data sharing cannot be avoided.

An appropriate cartographic design and layout is adapted for each spatial data scale representing different countries and island groups. Factors such as large extent of mapping area and huge space covered by ocean introduce a challenge for correct cartographical design. In addition, traditional mapping principles of design are dramatically changed in digital on-line Web cartography. The Development of a solution for a Web Map Atlas into server-client-geodatabase architecture in a constrained net environment requires us to investigate related technological problems including publishing of large size data files, data format of warehouse, access security, etc. Finally, the atlas is considered as a development methodology to build a "*Geographic platform*" as a step for GIS research and learning in the region.

Intergraph GIS suit such as GeoMedia was chosen to build an interactive environment for the new web-atlas. GeoMedia Web Map Professional and Web-Publisher are used for the atlas publishing, production, and design. The GIS Unit has received a GeoMedia software package as a part of its Registered Research Laboratory program membership, from Intergraph.

GeoMedia Web Map provides tools necessary to generate a map from the Web Map Server that can be provided to a client for viewing using a Web browser. Users on the client sites will be able to work with the map interactively and dynamically – zoom, pan, identify semantic attributes, change legend and scale, measure distances and areas, use the latest information from the data sources available etc. GeoMedia Web Map utilizes the ActiveCGM (Active Computer Graphics Metafile) format to generate client display and employs the technology of Active Server Pages (ASP) or java servlets for medium or thin clients. A map can be created from many different GIS data sources including binary spatially optimized format such as GeoMedia SmartStore.

This article describes solutions, which have been employed to achieve the some of the above listed objectives.

**Web Atlas and Distance Education in the Region**

One of primary goals of the Atlas design and the main target audience are respectively higher education and university students in the South Pacific. The USP is an international University serving 12 countries (Cook Islands, Fiji Islands, Kiribati Islands, Marshall Islands, Nauru, Niue, Tonga, Tuvalu, Tokelau, Solomon Islands, Vanuatu, and Western Samoa) where a majority of the students is studying by the Distance and Flexible Learning (DFL) mode.

There are a variety of Distance Learning tools such as classroom-based learning, visiting instructor-led training, videotape or satellite video-based training, and comp-aided E-learning from a live Internet connection, offline learning delivered by CD-ROM or via pre-download data, and mobile professionals or M-learning [4].

As an accumulator or focal point for knowledge, the Atlas should offer the user public data from different disciplines and for different disciplines at USP. The base is of course a set of GIS spatial data, but the geography, as a multi-disciplinary science, has provided the guidelines for the project. Summarized information in the forms of maps, spatial images, tables, texts, graphs and graphics from diverse cultures coming from Melanesia, Micronesia and Polynesia will be incorporated into the Atlas. From other perspective, this project is also seen as an important step for Information Technology and communication development in the region.

The ITC structure of USP in the region is designed in a hierarchical way. The main Laucala Bay Campus is in Fiji, another two are in Vanuatu and Samoa and there are fifteen (15) additional national and sub national USP centers, at least one in each USP member country. For the past ten years, the Geography Department at USP has included basic and intermediate GIS and introductory Remote Sensing (RS) courses within its undergraduate curriculum. As a result, more undergraduate students and regional GIS users have basic GIS/RS training. Today the GIS programme is expanding rapidly. Four courses are currently offered, two at 200-level (second year undergraduate) and two at 300-level (third year undergraduate). The number of students taking GIS courses has increased significantly during the last few years. The demands are greater than the Geography Department's GIS-Unit resources, largely because of a lack of staff and the number of workstations in the GIS/RS Laboratory. Classroom-based GIS/RS courses are offered in Laucala campus. Different types of DFL delivery are used for conducting GIS/RS courses for off campus students.

Classroom-based (summer schools) and satellite video-based training have been used for GIS DFL education in the Department. GIS/RS Summer schools are in high demand and popular throughout the region, but there are expensive and require additional staff members. Satellite video-based training is also expensive and more appropriate for lecture delivery. Learning from videotape suffers from lack of contact between a lecturer and students, and appropriate is mostly for lectures. E-learning is best for practical and laboratories in GIS DFL. It can also be used for theory learning.

Interactive Web Map Atlas is seen as being a critical component for comprehensive GIS and Geography E-learning as well as for other forms of DFL. It can serve many functions for DFL including delivering information and references for coursework and lectures and spatial and attribute data. At the same time, with regard to the limitation of network bandwidth of Internet and Intranet connections between the main campus and the remote-island centers, the development of a GIS multimedia support (CD-Rom) is necessary to address these constraints. GIS Atlas on CD-Rom may include a version of web-map interface and spatial warehouse for installation in the island Centers, and countries with rural and isolated communities, which have no Internet connections at all.

**Cartographical Issues: represent space ranging from an atoll less than a kilometer wide, to the "*USP Exclusive Economic Zone*" covering 13.5 million square kilometers**

There are a number of cartographical challenges designing and producing an Atlas for such a diverse region as the Southwest Pacific. Among of them are large variations in the areal sizes of the countries, the large extent of the mapping area, the time terrestrial object (date line) and the widely scattered and fragmented nature and extensive areas of ocean within the exclusive economic zones (EEZ) of most countries, differences in data sources availability, accuracy and timeless of data, map projection and datum, symbology and object classification, etc. The region is also the cradle of diverse cultures (Melanesian, Micronesian and Polynesian), which have to be depicted within the atlas content.

The project is to develop an interactive GIS web-atlas using spatial data range variation for twelve different Pacific Island countries. Table 1 highlights some of the differences between USP member countries' territories, which ranged from atolls of Tokelau with a total land area of 10 km$^2$ areas to the Solomon's Islands archipelago with an area of 28,450 km$^2$.

How to turn such scale differences into appropriate cartographic representation is a major challenge? There are currently no examples of Internet atlases at a large scale, between 1:50000 and 1:250000, which covers the islands countries of the Pacific.

It was decided to create one digital vector data set for each country and one separate data set for Exclusive Economic Zones (EEZ) that covers all 12 countries. The total size of each country varies significantly, and at the same time, taking into consideration the size of the terrestrial area for each territory, the base scales of mapping were defined as it is shown in Table 2. Base scales for individual countries were chosen mostly in accordance with the terrestrial sizes of main islands of the respective countries. The Pacific Region is mapped at the general-reference scale as an overview map and will have live links to individual countries for more detailed display and analysis.

Table 1: USP members countries

| Country | Capital | Population (est. 2003) | Area (sq km) | Coastline (km) |
|---|---|---|---|---|
| Cook Islands | Rarotonga | 21008 | 240 | 120 |
| Fiji Islands | Suva | 868531 | 18270 | 1129 |
| Kiribati | Bairiki | 98549 | 811 | 1143 |
| Marshall Islands | Majuro | 56429 | 182 | 370 |
| Nauru | Yaren | 12570 | 21 | 30 |
| Niue | Alofi | 2145 | 260 | 64 |
| Tokelau | None | 1418 | 10 | 101 |
| Tonga | Nuku'alofa | 108141 | 748 | 419 |
| Tuvalu | Funafuti | 11305 | 26 | 24 |
| Solomon Islands | Honiara | 509190 | 28450 | 5313 |
| Vanuatu | Port Vila | 199414 | 12200 | 2528 |
| Western Samoa | Apia | 178173 | 2944 | 403 |

Table 2: Scale definition for the countries

| Region | Scale |
|---|---|
| Pacific Region | 1 : 1,000,000 |
| Fiji Islands, Vanuatu, Solomon Islands and Western Samoa | 1 : 250,000 |
| Tonga, Tuvalu and Cook Islands | 1 : 100,000 |
| Nauru, Niue, Tokelau, Kiribati and Marshall Islands | 1 : 50,000 |

The digital spatial and attributive data set for each country will include not only topographical information, but also thematic data. Content of the thematic information will vary for each respective country. These variations are caused by differences in data availability and the cultural diversity of the countries. The priority for the first stage of Atlas creation is to design and publish general-referenced data, and later to add thematic layers and attributes for physical-geographical, environmental and socio-economical information.

The data that will be stored within a single digital spatial database for each country may include following:
  - General-referenced spatial and attributive data (coastlines, administrative boundaries, hydrology, coral reef cover, elevation, population sites, transportation etc);

- Physical-geographical and environmental spatial and attributive data (vegetation cover, soil types, climate characteristics, geology, protected areas, waste disposal sites etc);
- Social-economical spatial and attributive data based at national level on statistical databases from regional organization such as the Community of the South Pacific (SPC), the South Pacific for Applied Geosciences (SOPAC) and on USP members countries statistical offices for regional and country levels (demographic and population indicators, economic indicators, land use, health indicators, ethnic and religion, tourism indicators etc).

The regional data sets include generalized general-referenced spatial and attributive data (EEZ and politico-administrative boundaries, coastal lines, major population sites, shipping routes, etc) and layers with the main physical-geographical, environmental and social-economical indicators (basic demographic data, major economical and tourism indicators, main ecological problems faced by each country, etc). At the present, the work for data acquisition and processing, web pages design, Web Map Server set-up and testing has started. The implementation issues related data capture and preparation for the publishing is discussed in the next chapter. Questions related to Web Map publishing are outlined in the followed chapter.

**Implementation Issues**

*Spatial Data Capture*

Interrogation of spatial data availability in each country is one of the key elements for the Web-Atlas success. In each USP members' country, available spatial data may come from public and private sources. Taking into consideration the specific attributes of the Pacific Islands countries, the strategy of direct visiting to each target country for data acquisition was chosen. Preparation work for each visit includes contacting via emails respective USP centers for logistic support and managers and personnel in government land departments, environmental agencies, utilities and private organizations. Personal contacts with individuals in appropriate organizations have greatly facilitated the process. For each organization a letter of inquiry was prepared, which includes explanations of the project objectives and statement of the intended data usage, including the non-distribution of original vector data and the possibility of sharing the entire final data collection with each contributing organization.

Each country has several spatial data producers. Some digital data sets have been prepared with the help of international/regional consultants. As a rule, these organizations are preparing spatial data in different formats, with different specifications, spatial and attributive contents, coordinate systems, and quality requirements. These data are difficult to "normalize". The last factor forces project developers to *revisit* producer during the data processing. In the Pacific region, inter-country communications are very slow. Due to the cost factor, data acquisition, travels have been limited to one visit per country, thus limiting the effectiveness of data acquisition.

Sets of spatial and attribute data have already been gathered from six countries (Cook Islands, Marshall Islands, Kiribati Islands, Vanuatu, Nauru, and Fiji Islands) and have been used for compilation of the Atlas. About 90% of the data was acquired in digital form, mostly in the form of MapInfo table files or ArcView shape files formats, and as air photos and satellite images. For the moment, very few data have been collected in the form of topographic and thematic paper maps. For some countries data from different organizations are duplicated which provides the possibility of choosing data set with better quality. One of the project objectives was to collect accurate spatial data for building regional comprehensive spatial data warehouse. Most of digital vector data in visited Pacific countries came from the digitizing of paper maps compiled on the last quarter of twentieth century or even earlier with minor updates. Satellite images and aerial orthorectified photographs were used for the updating.

*Coordinate System Conversion*

Most of existing digital and paper maps in the South Pacific countries are complied in TM or UTM projections with custom parameters. There are some islands for which maps are available only in local coordinate systems. The same situation exists with the datum; which vary widely depending on the producers. Surveying geodetic controls are available for some islands; only some GCPs came from GPS surveying. It is normal to expect the poor positional accuracy for some data and it is beyond the scope of the project budget to conduct a full mapping cycle from the start.

A geographical or Latitude/Longitude coordinate system with the World Geodetic System 1984 (WGS84) datum was chosen for the Atlas spatial data storage. The longitude coordinate in the system extends from 0 to 360 degrees which allows the representation parts of some archipelagos, which extend across the International Date Line (180 degree East and West), for example Fiji group, to be depicted continuously in the same view. Coordinate system conversions of source data to Atlas warehouse storage include recomputation of coordinates from source projections and datum to Lat/Long and WGS84 for source spatial data with known coordinate systems; coordinate transformations from local grid to Lat/Long and WGS84 by using surveying control points and GCPs. For the web publishing, spatial data from the main warehouse has to be re-projected in TM projection and WGS84 datum and after be converted into GeoMedia SmartStore format.

*Spatial Data Input and Processing*

The methodology of creation of spatial data warehouse for the Atlas is to "normalize" or integrate existing data from different sources and minimize any fieldwork. Normalization implies conversion of data to a common coordinate system and vector format; topologic cleaning of data necessary, and usage of common specifications for spatial layers and attribute tables for all countries. As it was mentioned above, spatial data are acquired in different formats, often in MapInfo *table* file, ArcView *shape* files and Arc/Info coverages. Also some data comes in CAD formats as dxf/dwg and dgn. GeoMedia can read these formats directly for cartographic display and publishing, but the actual process consists of storing these data in a common format to organize unique data sets for each country as a unified warehouse. GeoMedia allows data to be stored in MS Access format for personal database. For the Atlas project, this format is convenient if we take into consideration relatively small volume of spatial data with a possibility to distribute this database via a CD-Rom to other USP Centers. It supports compound geometry, and MS Access is more or less fully functional Relational Database Management System (RDBMS).

The specifications were developed, which defines a list of database layers or feature classes and their geometry types, list of attributes for each feature tables and their data types, classifications for some attributes and key identification system. Processing of spatial data, before storing them in a country warehouse, include different stages:
  - Analysis of data quality; selection of the best data set from the collections, which come from different producers;
  - Format and coordinate system conversions;
  - Merging islands or separate data sheets in seamless spatial layers for each respective country; cleaning topology for separate layers and between related layers for congruence of spatial features.

Data, which are collected in the form of topographic and thematic paper maps, and airphotos and satellite images are scanned, geo-rectified and digitized for vector data input.

*Conversion to GeoMedia SmartStore Format*

speed bandwidth from/to USP. A solution to compensate for the low Internet connection is published data-sources to GeoMedia SmartStore format in GeoMedia Web Map Server. The SmartStore warehouse is a spatially indexed cache of 2D geometry and attributes used to facilitate the quick generation of maps without querying the original warehouse database [1, 2, 5]. There are some limitations to publishing in a SmartStore format, e.g., images cannot be published, spatial data has to be projected; attribute queries can be slow, incremental updates are not supported, etc.

Spatial data has to be is re-projected in TM projection and WGS84 datum from the source MS Access warehouses before converting to GeoMedia SmartStore format. The Atlas warehouse in MS Access format may contain more information and only some data, especially attributive, may be published in derived SmartStore warehouse. To propagate changes from MS Access warehouse to SmartStore warehouse, the old SmartStore cache has to be completely overwritten with the fresh data. GeoMedia SmartStore format is been using for the USP Web Atlas publishing.

**Web Publication Issues**

*Web Atlas Design in GeoMedia*

A classical geography atlas is an ordered collection of maps. The access to a country and/or thematic maps can be implemented as alphabetical order list, which contains a map title and respective page number. Special indexing usually includes spatial references to sites; each site has at least two entries – map title or page number and special index on the respective map.

An interactive web atlas is organized in a slightly different way. In the case of the South Pacific Atlas, there are 13 spatial warehouses, one for each of the 12 countries plus one for the region, and standard Web development tools used for implementation of navigation, search, cartographic visualization and analysis. The logic of the information search may be implemented similar as it was in a classical atlas. A search may be done by country name, theme for particular country, or site name.

The web site of the Atlas includes an opening page and introductory pages describing the Atlas, how to use it, an innovation that is offer, etc. Introductory pages have the entries to map displays. The navigation for map entries is organized in number of ways. Thus a user can go to the South Pacific regional interactive map page; turn on or off interesting regional map layers; link to thematic explanatory data; execute display and simple operations by applying the respective buttons or tools; apply hot-link tool and click on particular country zone to open pages with interactive maps for the country (Figure 1).

The Atlas pages have a frame with navigation aids, such as "Find the Country or/and Site", "Thematic Maps", and "List of Maps". "Find the Country or/and Site" entry link to the page where user can make a choice to open a page with a country or a site maps (Table 3) by marking of appropriate check box. The selection page also gives the choice to mark themes, which would be visualized within the map frame. Country map page can then be open with a zoom to the country of choice or it can be zoomed in to a selected site extent (Figure 2), and interesting layers will be checked. Pull down lists with check boxes, within Layer/Legend frame of country map pages; give the possibility to visualize appropriate reference or/and thematic layers. The layers are grouped in the classes (lists) such as general-reference maps, environment, climate, etc. The title of the particular Layer/Legend is linked to the explanatory page with information and metadata for the respective topic about the country or entire region. The explanatory pages contain textual, graph and tabular information with illustrations.

"Thematic Maps" entries have pull down hierarchical lists, which serve as links to thematic maps for a particular country of interest. The online user will display a thematic country map page with requested layers in vector data formats. "List of Maps", the entry that leads to the page, lists all available maps from the Atlas warehouses for different countries with available layers. These layers are stored in GIS vector format. Other geographical, historical, and sociological information such as digital imagery, movies, sounds, etc, will be incorporated in a future development of the Atlas.

Table 3: Web Atlas entries for Sites

| Country | Site | Country | Site |
|---|---|---|---|
| Cook Islands | • Northern Group <br> • Southern Group <br> • Rarotonga | Kiribati | • Gilbert Islands <br> • Line Islands <br> • Phoenix Islands |
| Samoa | • Upolu <br> • Savaii | Tokelau | • One entry for the entire country |
| Fiji Islands | • Viti Levu <br> • Vanua Levu / Taveuni <br> • Yassawa / Mamanucas <br> • Lomaiviti Group <br> • Lau group <br> • Kadavu group <br> • Rotuma | Vanuatu | • Efate <br> • Tafea <br> • Shepherds <br> • Epi <br> • Paama <br> • Ambrym <br> • Pentecost <br> • Malakula <br> • Ambae-Maewo <br> • Santo-Malo <br> • Banks-Torres |
| Marshall | • One entry for the entire country | Tuvalu | • One entry for the entire country |
| Solomon Islands | • Temotu <br> • Makira-Ulawa <br> • Malaita <br> • Guadalcanal / Central <br> • Isabel <br> • Western <br> • Choiseul | Tonga | • Vavau group <br> • Haapai group <br> • Tongatapu / Ata |
| Nauru | • One entry for the entire country | Niue | • One entry for the entire country |

The map design of the Atlas is done in GeoMedia Professional; standard styles from GeoMedia symbol library can be used or/and new symbol files can be created using Define Symbol File toolbox. As was mentioned above, GeoMedia Web Map Server is used for publishing maps within the Atlas. GeoMedia Web Map employs the technology of Active Server Pages or java servlets (depending on the Web publishing software) to generate an interactive map on a client browser. In addition to the GeoMedia Web Map tools, Web page designers have all the standard tools available to enhance the authoring Web page using standard Web development content, such as HTML, Dynamic HTML, XML, VBScript, JScript, JavaScript, and other Active X controls [3].

GeoMedia Web Map Publisher is deployed to set up a primary web site for each country or regional map. After that, standard web development tools can be used to enhance the Atlas pages. The GeoMedia Web Map Publisher may be used for web site maintenance [2].

*Development of a Multimedia Support for the USP Centers*

One of critical component of the project is to make the atlas and associated data available to all USP Centers and members' countries. Intranet/Internet connection and videoconference are provided from main USP Laucala Campus to other Centers through a satellite network. The nominal bandwidth of this network is 256 kbps, but practically it is much slower, especially for some countries. In addition, some USP Centers are still using very old computer hardware; sometimes with CPU speed of around 200 MHz etc. There is a significant gap between Fiji and the other remote Pacific islands countries in terms of accessibility to communication media.

Practically, many USP Centers do not have reliable Intranet/Internet connection for E-learning. This will affect the access to the Web Atlas as well. Such situations require the development of CD-ROM version of the Atlas for local installations.

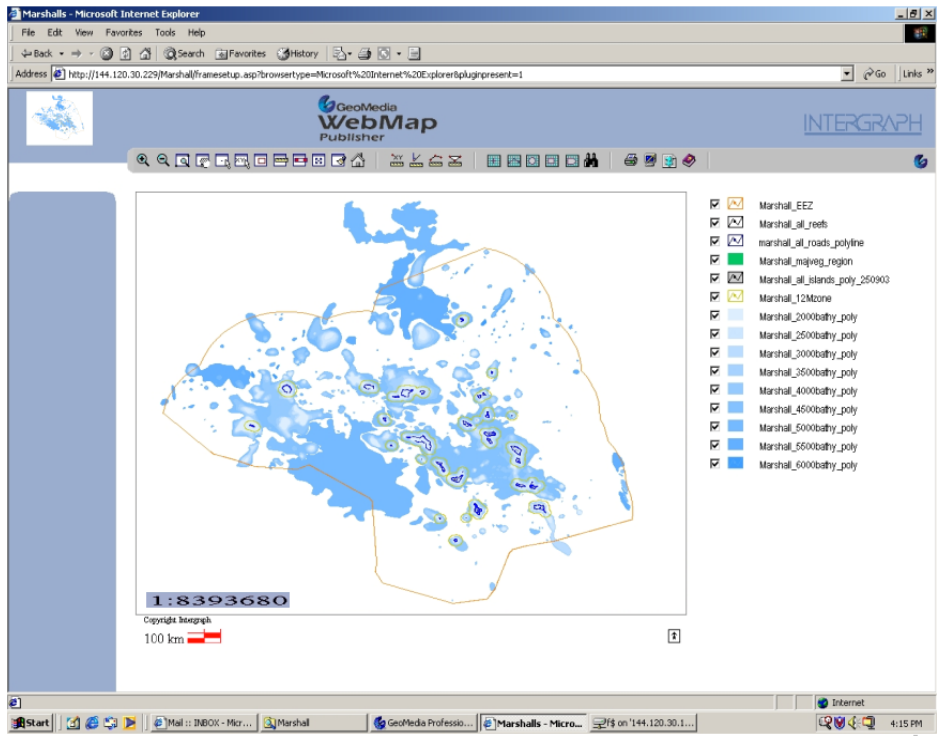

Figure 1: Under construction page with interactive maps for the Marshall Islands group, (the page is under construction)

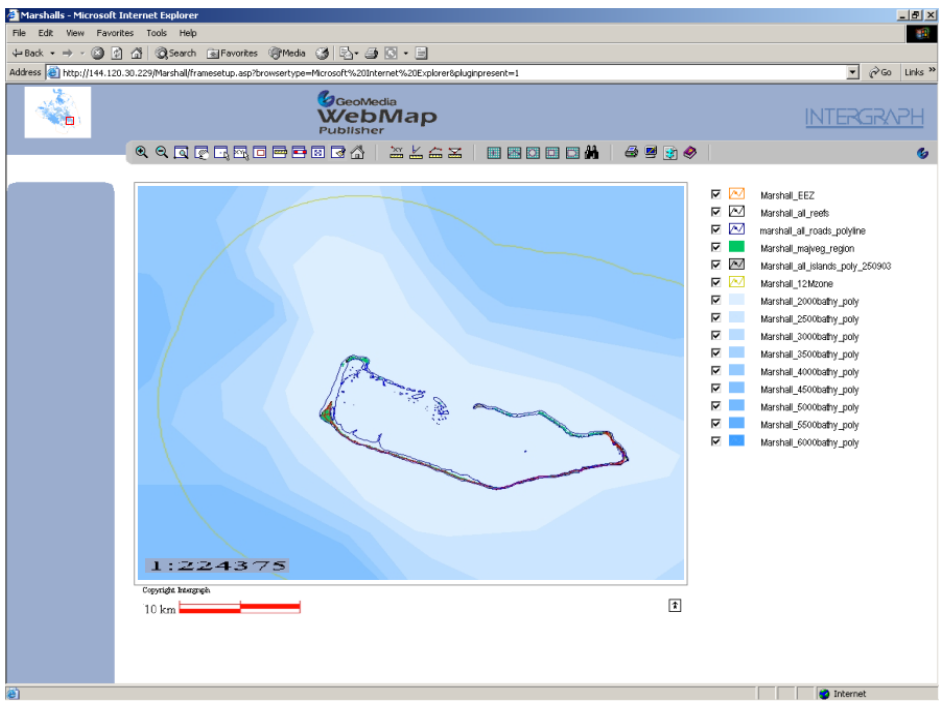

Figure 2: Interactive maps with the zoom in Majuro Atoll site, Marshall Islands group, (the page is under construction)

**Conclusion**

This article outlines first stage of development of the interactive South Pacific Web Atlas. So far, six spatial warehouses for the countries have been created, as well as a warehouse for the region.

These warehouses contain general-reference and environmental layers. The Web Atlas site is still in the development and design stage. The Atlas will be modified, as more information becomes available. A number of staff members and students from the USP GIS Unit, Geography and Computer Science Departments are taking part in this project.

Research and methodology development in that direction remains a priority to enhance further the USP-based GIS E-learning program and to extend its usefulness to the wider GIS user community in the USP region and beyond. The Web Atlas can facilitate, on a truly regional scale, the sharing of information required for research and education, and can serve as an "engine" for new GIS partnerships in the Pacific.